\documentclass[]{spie}  

\usepackage{amsmath,amsfonts,amssymb}
\usepackage{graphicx}
\usepackage[colorlinks=true, allcolors=blue]{hyperref}
\newcommand{\masyr}{${\rm mas\, yr^{-1}}$}

\title{High-precision astrometry towards ELTs}

\author[a,b]{Davide Massari}
\author[a]{Giuliana Fiorentino}
\author[b]{Eline Tolstoy}
\author[c]{Alan McConnachie}
\author[d]{Remko Stuik}
\author[a]{Laura Schreiber}
\author[c]{David Andersen}
\author[e]{Yann Cl\'{e}net}
\author[f]{Richard Davies}
\author[e]{Damien Gratadour}
\author[d]{Konrad Kuijken}
\author[g]{Ramon Navarro}
\author[h]{J\"{o}rg-Uwe Pott}
\author[h]{Gabriele Rodeghiero}
\author[i]{Paolo Turri}
\author[b]{Gijs Verdoes Kleijn}

\affil[a]{INAF-Osservatorio Astronomico di Bologna, via Ranzani 1, 40127, Bologna, Italy}
\affil[b]{University of Groningen, Kapteyn Astronomical Institute, NL-9747 AD Groningen, The Netherlands}
\affil[c]{Herzberg Astronomy and Astrophysics, National Research Council Canada, 5071 West Saanich Road, Victoria, BC V9E 2E7, Canada}
\affil[d]{Leiden Observatory, Leiden University, Postbus 9513, 2300 RA Leiden, The Netherlands}
\affil[e]{LESIA, Observatoire de Paris, CNRS, UPMC, Universit\'{e} Paris-Diderot, 5 place Jules Janssen, 92195 Meudon, France}
\affil[f]{Max-Planck-Institut f\"{u}r extraterrestrische Physik, Postfach 1312, D-85741, Garching, Germany}
\affil[g]{NOVA Optical IR Instrumentation Group, P.O. Box 2, 7990 AA Dwingeloo, The Netherlands}
\affil[h]{Max-Planck-Institut f\"{u}r Astronomie, K\"{o}nigstuhl 17, D-69117 Heidelberg, Germany}
\affil[i]{Department of Physics and Astronomy, University of Victoria, 3800 Finnerty Road, Victoria, BC V8P 5C2, Canada}
\authorinfo{Further author information: (Send correspondence to D.M.)\\D.M.: E-mail: davide.massari@oabo.inaf.it, Telephone: +39 051 2095318\\  G.F.: E-mail: giuliana.fiorentino@oabo.inaf.it, Telephone: +39 051 2095318}

\begin{document} 
\maketitle

\begin{abstract}
With the aim of paving the road for future accurate astrometry with MICADO at the European-ELT, we performed
an astrometric study using two different but complementary approaches to investigate two critical components
that contribute to the total astrometric accuracy.
First, we tested the predicted improvement in the astrometric measurements with the use of an 
atmospheric dispersion corrector (ADC) by simulating realistic images of a crowded Galactic globular cluster. 
We found that the positional measurement accuracy should be improved by up to $\sim2$ mas with the ADC,
making this component fundamental for high-precision astrometry. 
Second, we analysed observations of a globular cluster taken with the only currently available Multi-Conjugate 
Adaptive Optics assisted camera, GeMS/GSAOI at Gemini South. Making use of previously measured proper 
motions of stars in the field of view,
we were able to model the distortions affecting the stellar positions. We found that they can be as large as
$\sim 200$ mas, and that our best model corrects them to an accuracy of $\sim1$ mas. We conclude that future 
astrometric studies with MICADO requires both an ADC and an 
accurate modelling of distortions to the field of view, either through an a-priori calibration or an 
a-posteriori correction.

\end{abstract}


\section{INTRODUCTION}
\label{sec:intro}  

Accurate astrometry is one of the major drivers for diffraction limited Extremely Large Telescopes (ELTs). 
To reach diffraction limited observations, the Multi-AO Imaging Camera for Deep Observations (MICADO), one of the 
first light instruments for the European-ELT,
will be assisted by an Adaptive Optics module (MAORY, \cite{diolaiti10}) providing both a Single Conjugate 
(developed jointly with the MICADO consortium) and a Multi Conjugate modes.
The goal of MICADO is a relative astrometric accuracy for bright and isolated stars of 50 $\mu$as over a central,
circular field of 20 arcsec diamater.  
To determine if such an ambitious goal is feasible, a dual approach is taken.
To simulate stellar fields as they would be seen by the SCAO with the predicted instrumental Point Spread Function (PSF) and 
analyse the resulting realistic images will test the predicted performance, and help to optimise the 
instrumental design. 
In addition, present-day astrometric studies with existing MCAO facilities are crucial 
to test the main sources of inaccuracy not related to the specific instrumental design or telescope. 

In this paper we present both these approaches, as complementary studies. 
In particular, we start by investigating in Section \ref{sim} how to best reach the astrometric requirement 
for MICADO by quantifying the errors associated to one of the most important components in the light path: 
the atmospheric distortion corrector (ADC). This investigation is carried out making simulations with the 
SCAO module PSF of the central region of a crowded globular cluster, for a field of view of $2 \times 2$ arcsec.
This size is small enough for PSF variations not to be important, but big enough to contain a sufficient 
number of stars.

Then, in Section \ref{real}, we also present the results of an astrometric study performed with Gemini Multi-Conjugate Adaptive Optics 
System (GeMS) observations of the Galactic globular cluster NGC6681. This cluster is the most centrally concentrated in the 
Galaxy, and thus represents a major observational challenge in terms of stellar crowding. We have previously determined 
proper motions by comparing two Hubble Space Telescope epochs (\cite{massari13}). This makes this cluster an ideal 
candidate to test the effects that will be introduced by MCAO corrections on proper motion measurements 
and related uncertainties. 
In particular, we looked for systematic distortions introduced in the GeMS images by observing through 
both J and Ks filters and we quantify their impact on the astrometry. 
Though previous studies have already tested the astrometric performance of MCAO cameras (e.g. \cite{neichel14a,
ammons14, lu14, fritz16}), our investigation is the 
first to address the detailed structure and sizes of distortions in a MCAO instrument and will 
therefore be the starting point for understanding MCAO astrometry capabilities and any future improvements in 
the calibration and data reduction strategy for ELT observations.

\section{MICADO ADC simulations}\label{sim}

One of the most severe observational issues concerning accurate astrometry is {\it atmospheric dispersion}. 
Since the refractive index of the atmosphere, $n$, depends on the wavelength,
the observed angular distance between two sources is altered depending on the difference of their colours.
Moreover, atmospheric dispersion elongates the shape of the PSF along the zenith, thus 
reducing the Strehl ratio and affecting the precision of any astrometric measurement.
This is especially true in crowded fields, where the position of faint sources is affected by 
the broader wings of the PSF of bright neighbours.
Since the combined effect can have an impact as large as few mas on the astrometry (\cite{trippe10}), an adequate
correction is mandatory. Previous dedicated studies determined as best solution for MICADO observations to
introduce a counter-rotation based ADC located at the pupil of the instrument. 
A detailed description of such a component is provided in an internal communication document by Remko Stuik.

\begin{figure*} [ht]
    \begin{center}
    \includegraphics[width=8.5cm]{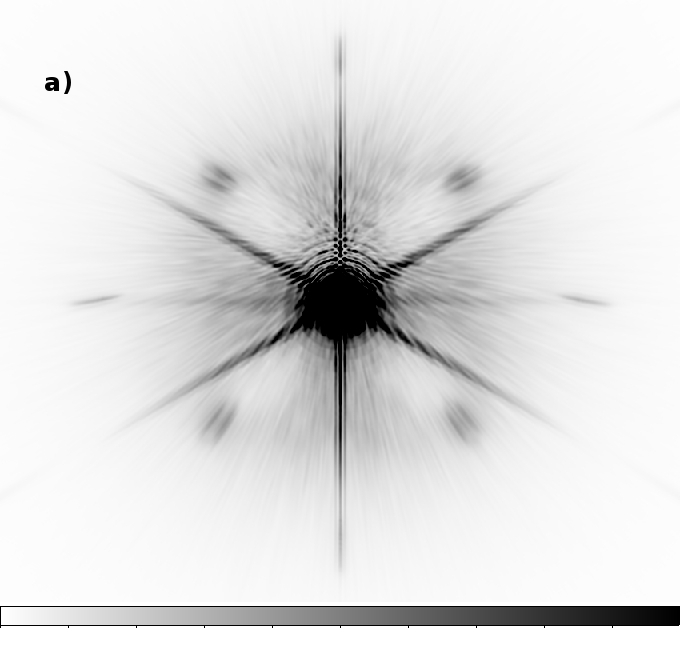}
    \includegraphics[width=8.5cm]{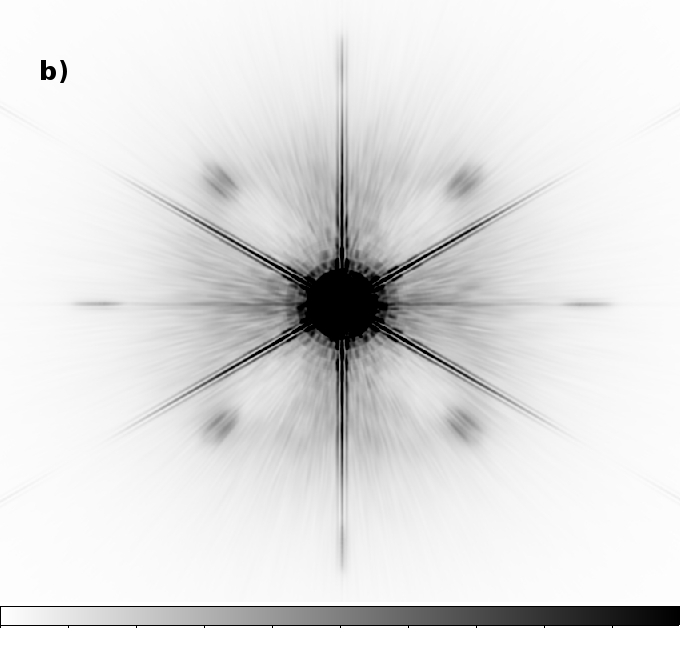}
        \caption{\small Comparison between the PSFs predicted for a 20 s MICADO exposure without (Figure \ref{psf}a) 
        and with (Figure \ref{psf}b) the inclusion of an ADC in the light path. Each image has a size of 
        $4096\times4096$ pixels and is orientated with the X and Y axis corresponding to Right Ascension
        and Declination, respectively.}
        \label{psf}
    \end{center}
\end{figure*}

In order to quantify the impact of using an ADC on the astrometric performances of MICADO, we simulated realistic 
images of a Galactic globular cluster field using the predicted instrumental PSF of the SCAO module. 
The PSF were generated from a preliminary set of AO simulations on the COMPASS platform by Yann Clenet 
(\cite{clenet13}). The telescope spiders and segmentation 
were added to the individual frames and the PSF was shifted based on wavelength and zenith angle. 
An optical ZEMAX model was used to compute the correction as a function of wavelength for an ideal ADC 
with counter-rotating double prisms, assuming a fixed zenith angle of $60$ degrees. 
We stress that this is not a full end-to-end model, and 
that the interactions between all of the various sources of error, as well as optical and alignment errors, 
are not fully included.  
The PSFs were computed on axis, in the standard Ks-band filter, with an atmospheric coherence length r$_{0}=0.129$ m
at a wavelength of 0.5 $\mu$m and a resulting Strehl ratio of $0.76$. 
The AO simulations ran for 1 hour, with a frame rate of 1 kHz, but only one of
every 1000 PSF images was saved. Therefore, each PSF of 1 ms represents 1 s of data. 
To perform our astrometric tests,
we summed $N$ independent PSF realisations to produce exposures of $N$ seconds.
Although the global PSF is a correct representation of the stated exposure time, 
the noise statistics are not. 
The PSF predicted for a 20 s MICADO exposure without (a-panel)
and with the inclusion of the ADC in the light path (b-panel) are shown in Figure \ref{psf}.

It is already clear that the ADC can correct the PSF, making it more symmetric
and removing many of the speckles near the upper region of the PSF core. 
The PSF also appears sharper, with a Strehl ratio that increases from $0.35$
to $0.76$.
In the following we will quantify the astrometric improvement due to the introduction of the ADC, 
demonstrating how fundamental this component is for astrometric studies of crowded stellar fields 
with MICADO.

\subsection{Input catalogue and Simulations}

To carry out our investigation, we simulate a realistic astrophysical problem.
A natural choice is the crowded stellar field of a Galactic globular cluster (GC). In fact, GCs
have routinely been the subject of detailed astrometric studies (e.g. \cite{bellini14, watkins15}), 
and because of the availability of bright guide stars, 
they are ideally suited to be studied with diffraction-limited AO observations. 
In our investigation we want to simulate only a small region of the sky,
where the PSF can reasonably be assumed to be constant across the entire field of view (FoV).
Several previous studies have demonstrated that the PSF in AO images varies in a way 
that is very difficult to predict (see e.g. \cite{fiorentino14, saracino15, turri15, massari16}).
However, the introduction of a variable PSF is beyond the current scope of this work and will
be addressed in future investigations.
At the same time, we need to simulate a number of stars large enough to draw 
statistically significant conclusions. For these reasons, we choose to simulate the innermost
2 arcsec $\times$ 2 arcsec region of the crowded core of the GC M3 (\cite{massari16b}). 
Since M3 is known to have a flat density distribution in its core (\cite{miocchi13}), 
it is correct to assume that stars are uniformly distributed in the central
2 arcsec $\times$ 2 arcsec region, and we can build the input catalogue simply 
by distributing the stellar positions randomly.

We determined the realistic number of stars to be simulated using the Hubble Space Telescope (HST) 
catalogue of M3 (\cite{anderson08}). 
After correcting it for incompleteness effect, we found the total number of stars in our FoV to be $\sim650$.
This is the population we will use to create a realistic simulation of a MICADO image of M3.
We also reproduced a realistic distribution of stellar magnitudes by using the theoretical models 
taken from the Basti archive (\cite{pietrinferni06}). 

\begin{figure*} [ht]
    \begin{center}
    \includegraphics[width=\columnwidth]{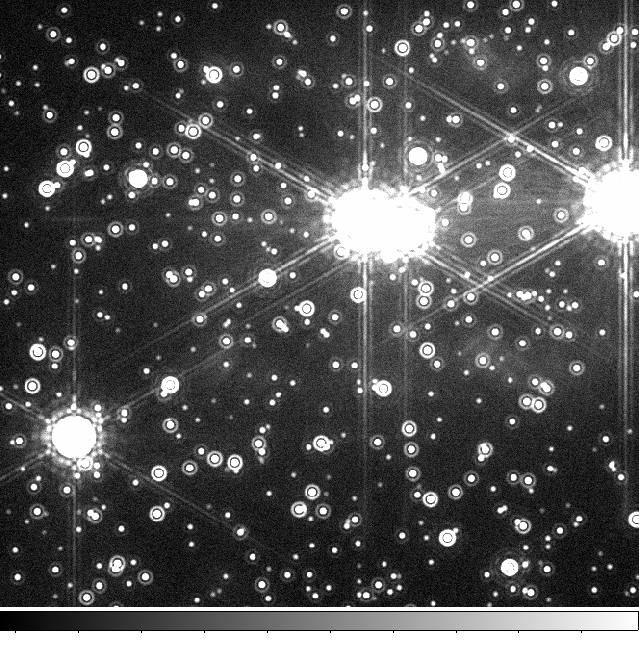}
        \caption{\small Example of one of the simulated stellar field analysed in this work. The image simulates a 20
        s exposure of the inner $2\times2$ arcsec of the GC M3 taken in the Ks band with MICADO, with the inclusion of
        the ADC.}
        \label{clu}
    \end{center}
\end{figure*}

The software used to create the simulated images is described in detail in \cite{deep11}.
A few of the technical specifications used in that paper were updated for this investigation.
In particular, we now use a primary mirror with an outer diameter of 37.0 m, an 11.1 m internal diameter, 
6 spiders of 40 cm width every 60 degrees, for a total effective area of 947.3 m$^{2}$.
Given a pixel-scale of 3 mas/pixel\footnote{Note that the current predicted diamater of the E-ELT is $38.5$ m, 
while the MICADO pixel-scale is currently set to 1.5 mas/pixel for the high spatial resolution mode.} 
our $2\times2$ arcsec images have a size of $667\times667$ pixels.
We created two sets of 
simulated images, with and without the inclusion of the ADC, with exposure times of 2 s, 4 s, 5 s, 10 s, 20 s and 120 s. 
These images were built following a random dither pattern to reproduce a typical scientific observation.
An example of an image simulated in this way and using a 20 s exposure PSF with ADC is shown in Figure \ref{clu}.

\subsection{Astrometric analysis}

The source detection and extraction in the simulated images has been carried out with the DAOPHOT (\cite{stetson87}) suite
of software. The PSF model used to fit the light profile of each star was determined
on the images, without exploiting any a-priori knowledge. This is to accurately mimic what happens routinely 
when dealing with real imaging data. 
Here, the best solution turned out to be a Moffat function. No degrees of spatial variation
were necessary, since by construction the PSF does not vary across the images FoV. This model was
then fit to all of the sources above a 3$\sigma$ threshold of the sky background by ALLSTAR, to give 
a catalogue of stellar positions and instrumental magnitudes as output.

\begin{figure*} [ht]
    \begin{center}
    \includegraphics[width=\columnwidth]{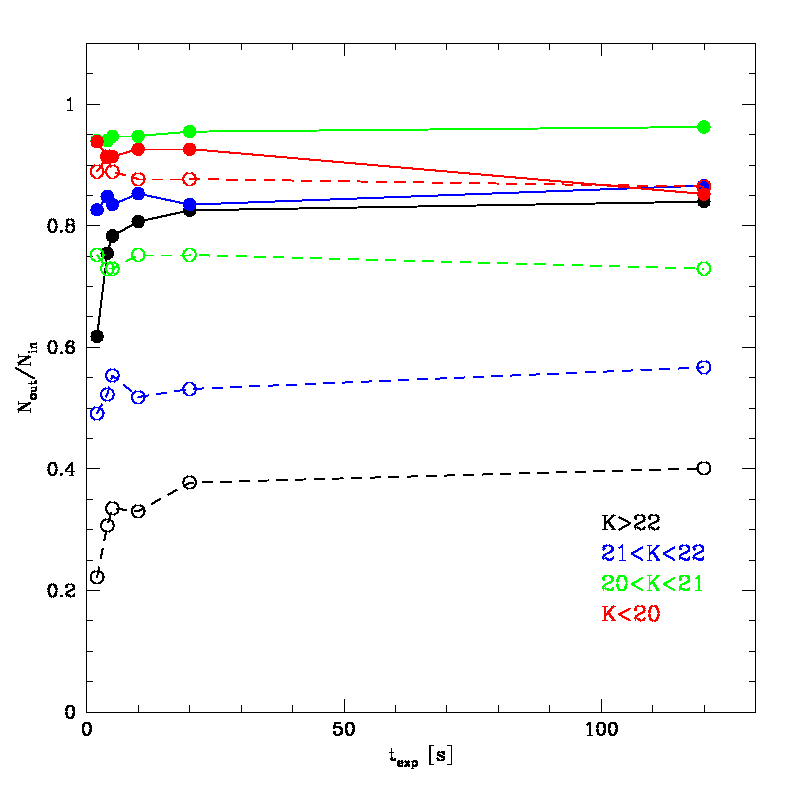}
        \caption{\small Completeness as a function of exposure time and stellar magnitude. Filled symbols
        connected by solid lines mark the results for the simulations including the ADC. Empty symbols
        connected by dashed lines those for the simulations without the ADC. Different bins of magnitude
        are shown with different colours, as explained in the labels.}
        \label{complet}
    \end{center}
\end{figure*}

The first interesting
result comes from a comparison of the number of input sources with the number of sources actually recovered
in the analysis, that is a sort of completeness test (see also \cite{deep11, greggio12} for other
detailed analysis on the achievable completeness with simulated MICADO data). The achieved completeness 
is shown in Figure \ref{complet}. It does not take into account false detections, which are stars found
that were not in the input list, and were discarded by cross-matching input and output catalogues.

\begin{figure*} [ht]
    \begin{center}
    \includegraphics[width=\columnwidth]{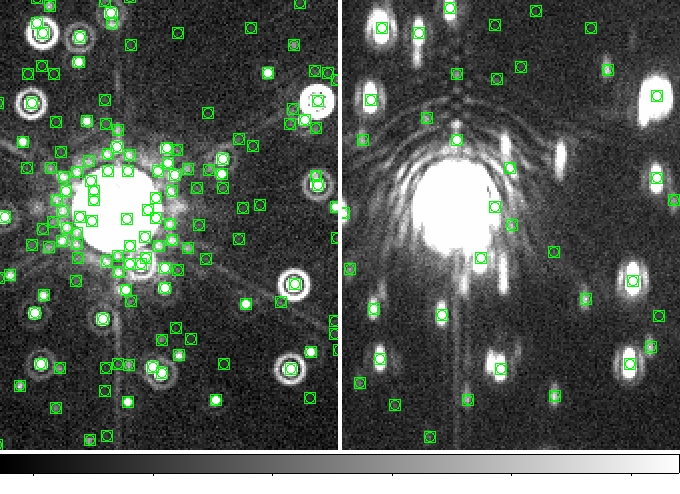}
        \caption{\small $190\times230$ pixels zoomed region of one of the simulated images. Left panel shows the 
        simulation with the inclusion of
        the ADC, and vice versa. Green symbols show the sources recovered by the reduction software. 
        For all these simulations, the direction of the PSF 
        distortion coincides with the Y component of the detector.}
        \label{confr_compl}
    \end{center}
\end{figure*}

As expected due to the superior PSF quality, the performance obtained in the ADC-case is strikingly better than that without the ADC. 
This shown in Figure \ref{confr_compl}. In the ADC-case (left panel of Figure \ref{confr_compl}) the software is able to pick up most
of the true sources (plus some PSF artefacts that however can be easily identified and discarded because of 
their non-stellar shape). In contrast,
in the no-ADC case (right panel of Figure \ref{confr_compl}) the elongated PSF causes the fainter stars to fall below the detection limit,
while bright stars with bright companions can often no longer be recognised as independent sources.
This is already an important indication of how important it will be to have an ADC assisting MICADO at the E-ELT,
not only for astrometry but also for purely photometric purposes.

\begin{figure*} [ht]
    \begin{center}
    \includegraphics[width=\columnwidth]{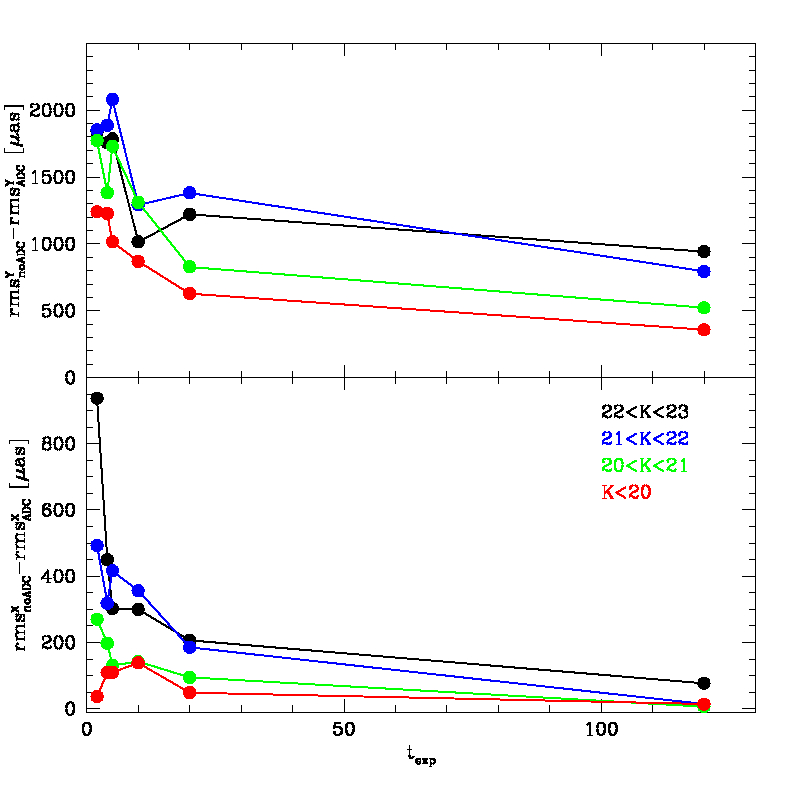}
        \caption{\small Difference between the astrometric performance between the no-ADC and the ADC case as a function of exposure time
        and stellar magnitude (see the labels). Symbols are colour-coded as in Figure \ref{complet}. The larger difference
        found in the Y-direction of the detector is due to the direction of the PSF distortion in our simulations.}
        \label{test1}
    \end{center}
\end{figure*}

We tested the astrometric performance by comparing the input positions with those
recovered by the software as output. In particular, we considered all of the stars in a given exposure,
and computed the root mean square (rms) of the difference between input and output positions. Then, we divided our sample
of stars in bins of magnitude and computed the corresponding mean rms value. Magnitude bins were defined
in order to have a statistically significant number of stars (at least 70) in each of them. Finally, we determined the
difference of the rms values between the non-ADC and the ADC case, and we show its behaviour
with respect to exposure time and input magnitude in the two panels of Figure \ref{test1}.
The X- and Y- direction of the detector have been analysed separately.

The improved performance with the inclusion of the ADC is marked. This is especially
evident in the Y-component (upper panel of Figure \ref{test1}), since it is the direction where the 
atmospheric dispersion most affects the PSF shape in our simulations because of our chosen field orientation. 
The difference in the performances in 
the X-direction is less conspicuous, but clearly improving for fainter magnitudes and shorter exposure times. 
Of course for real observations without the ADC, such a distortion usually has a significant component in both the 
directions depending on the orientation, thus affecting strongly the achievable astrometric accuracy in both axes. 
When combining in quadrature the improvement in both components, our tests reveal that the minimum improvement
(obtained with the brightest stars in the longest exposure) amounts to $\sim370~\mu$as, while the maximum
value (for the faintest stars in the shortest exposure) amounts to $\sim2000~\mu$as, in fair agreement
with the prediction of \cite{trippe10}.
Therefore, under the hypothesis that the PSFs used are realistically reproducing the reliable effect of the 
ADC, our study strongly supports the need of a high-functioning ADC in MICADO to achieve accurate astrometry.

\subsection{Discussion on the astrometric accuracy}

As stated in the Introduction, MICADO has the ambitious goal of reaching a relative astrometric accuracy of $50~\mu$as
for bright and isolated stars.
To reach such a goal, all of the components of the instrumental design have to be carefully tested in order to
minimise the contribution of any systematic source of astrometric error to the overall budget.
In this investigation, we focused on the impact that the ADC has on the SCAO performance of  
MICADO in simplified conditions, namely that the simulated FoV is small and on axis, and without considering 
the effects of the camera distortions, but for a realistic observational setup and science case.
Our findings clearly demonstrate that the inclusion of the ADC is a fundamental requirement
to reach high-precision astrometry, since in the simulated conditions it significantly improves the astrometric 
performance by at least $\sim370~\mu$as.

One of the future goals we will pursue is to improve the PSF generation, including the realistic ADC manufacturing 
errors together with the instrumental ones. 
Another necessary future step is to accurately quantify the total contribution of the ADC 
to the overall astrometric error budget. To do so, we also need to quantify the contribution to the error coming
from the inability of current software in modelling the PSF.
Starfinder (\cite{diolaiti2000}) is the ideal software to do this, as it is able either to model the PSF directly on the single 
frames, or to use as input the same PSF used to simulate the images, thus bringing to zero the uncertainty due to the PSF modelling. 
However, we stress that the procedure we followed in this study is routinely and necessarily 
followed when analysing real images given that no sufficiently accurate a-priori knowledge of the PSF is usually
available, the only exception 
being PSF reconstruction experiments (e.g. \cite{jolissaint12}).
Once the intrinsic contribution to the astrometric error budget from the optical design, including the ADC, 
is estimated, we will also be able to quantify the potential effects of the reduction software.
This will be extremely important in order to determine what are the likely future software requirements,
and to test how PSF reconstruction techniques could reduce the overall astrometric uncertainty budget.

\section{Real data}\label{real}

The practical requirement to measure stellar proper motions (PMs) is to determine the displacement of stellar positions between two
(or more) epochs. However, the physical motion of a star is not the only contribution to such a displacement.
In fact, any effect artificially altering the observed position of a star with respect to the true one introduces a 
\textquotedblleft distortion\textquotedblright that, without a proper treatment, is degenerate with the PM signal.
For this reason it is fundamental to disentangle the effect of such distortions before any astrometric investigation.

In this respect, MCAO systems are particularly complex to deal with. In fact, deformable mirrors conjugated to high
altitude layers far away from the pupil can induce field distortions that significantly affect the overall astrometric accuracy
(see e.g. the discussion in \cite{neichel14a}).
Since the magnitude and structure of distortions in MCAO data might change with differing seeing conditions, asterisms, airmass 
(\cite{rigaut12, lu14}) and are as yet poorly investigated, 
our ability to determine how accurately they can be corrected remains uncertain.
Calibrating the camera distortions and applying the general solution to any data-set is unlikely to prove 
sufficiently accurate for high-precision astrometric studies, because the distortions change from case to case.
One of the possible solution to this problem, would be to correct each exposure of a data-set with its own absolute solution, 
but this requires the availability of a distortion-free reference to break the PM-distortion degeneracy.

In the meantime a case where we can still make an accurate assessment of these effects is the Galactic GC NGC6681. 
This is because, both the distortion free positions in a past epoch and the PMs of the stars in the cluster FoV are known from Hubble Space Telescope (HST) 
measurements (\cite{massari13}, hereafter Ma13). Recent observations have been taken with the MCAO camera GeMS 
(\cite{neichel14b}) for this cluster (Programme IDs: GS-2012B-SV-406, 
GS-2013A-Q-16, GS-2013B-Q-55, PI: McConnachie). Therefore, in this case we have all of the necessary ingredients to determine 
the distortions caused by the MCAO on the GeMS camera.

\subsection{The method}

Because we have distortion-free stellar positions at the epoch of the first HST measurement (GO:10775, PI:Sarajedini)
and their PMs from subsequent HST epochs (Ma13), it is possible to build a distortion-free reference frame at the 
epoch of GeMS observations. In this way,
the differences between the observed GeMS positions and those from HST projected at the GeMS epoch are only
due to distortion terms.

In order to be as accurate as possible, only NGC6681 stars with a PM uncertainty smaller than 0.03 mas yr$^{-1}$
were used to build the reference frame. Their large number ($7770$) allow us to accurately sample
the area in common between the HST and the GeMS data sets.

The distortion-free reference at the GeMS epoch were aligned in Right Ascension (RA) and Declination
(Dec), and then all the GeMS exposures were registered to this reference frame to estimate their distortion maps.
The GeMS data set is composed of $8\times160$ s exposures in both the J and the Ks filters,
dithered by a few, non-integer pixel steps to cover the intra-chip gaps of the camera and
to allow a better modelling of the PSF.
All of the details concerning the reduction of the images will be described in a forthcoming paper
(Massari et al. in prep.). Briefly, stellar raw positions (x$_{i}^{r}$, y$_{i}^{r}$) were obtained via 
PSF fitting using the DAOPHOT suite of software (\cite{stetson87}) and following the procedure 
described in \cite{massari16}. Each of the four chips of the camera was treated separately. 
The PSF best-modelling was achieved by fitting the light profile of several hundred
bright, isolated stars with a Moffat function and allowing the fitting residuals 
to be described with a look-up table that varies cubically across the FoV.
By matching each exposure raw positions to the distortion-free reference using a 5th-order polynomial,
the corrected, distortion-free GeMS positions (x$_{i}^{c}$, y$_{i}^{c}$) were obtained.

The 5th-order polynomial turned out to be the best choice when trying to balance
the improvement of the rms of the transformations. This was determined by attempting to keep 
the order as low as possible, so as not to introduce spurious effects due to excessive degrees of freedom.
In particular, when moving from the third order to the fourth and fifth, the rms of the transformations 
improved by $\sim4$\% per order, leading to a final rms of $\sim1$ mas. Instead, for the following orders the improvement 
was only by $\sim1$\%.

\subsection{Quantifying GeMS distortions}

\begin{figure*} [ht]
    \begin{center}
    \includegraphics[width=8.5cm]{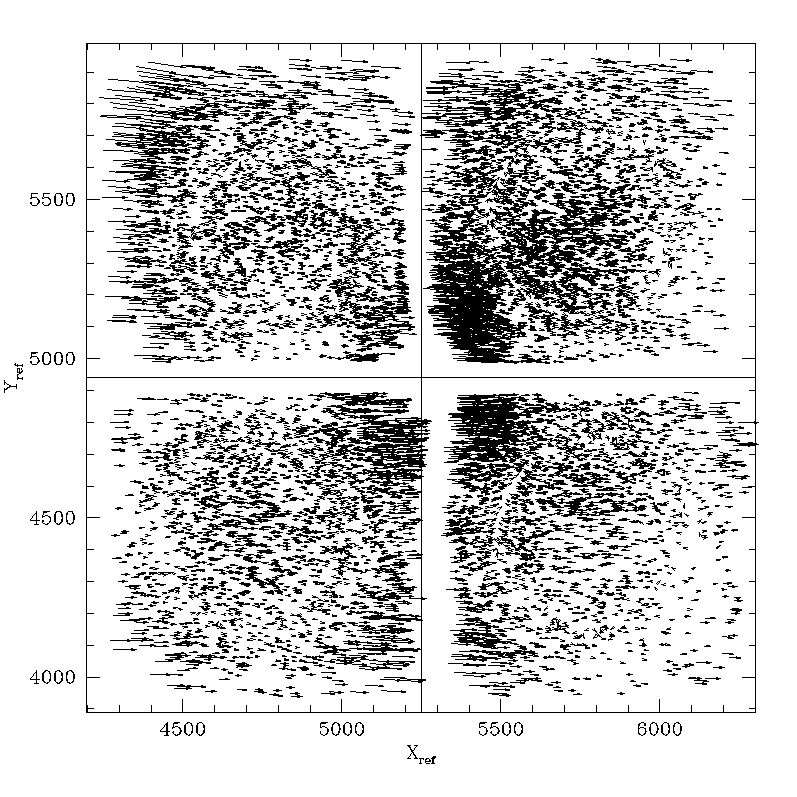}
    \includegraphics[width=8.5cm]{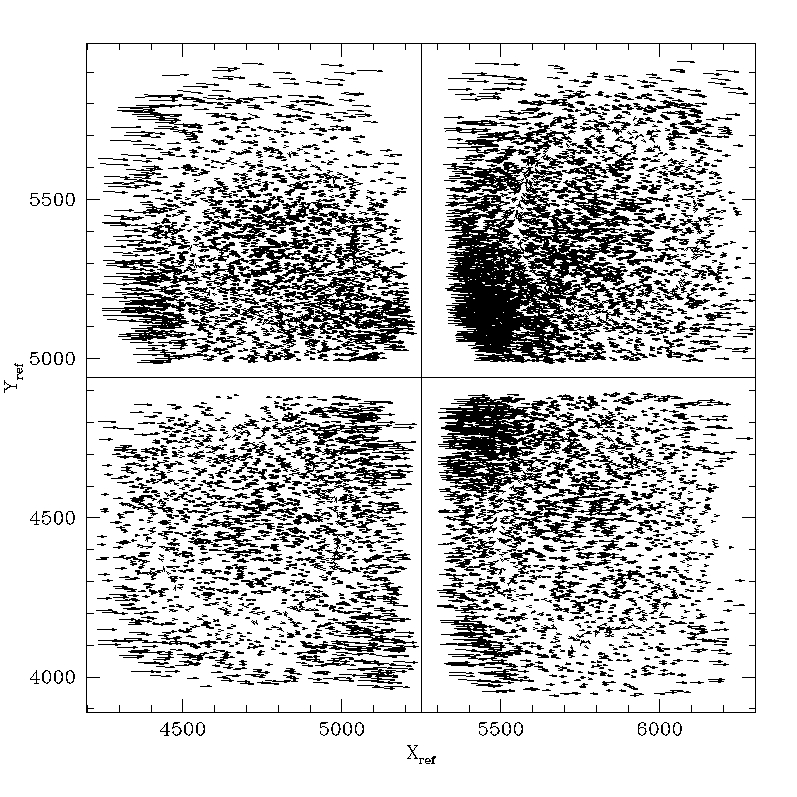}
        \caption{\small Average distortion maps for the J filter (left panel) and Ks filter (right panel).
        Vectors length is multiplied by a factor of 20 to highlight the distortion structures.}
        \label{dmaps}
    \end{center}
\end{figure*}

The aim of this analysis was to quantify the distortions that affect GeMS astrometry. 
We stress that the contribution to these distortions
does not come only from the instrumental geometric distortion, but also from all of the effects that
artificially alter the position
of a star such as anisoplanatism effects or an imperfect PSF modelling, and thus affect our ability to measure 
the true PM of that star.
The average GeMS distortion maps for the J and Ks images are shown in Figure \ref{dmaps}. 
Each single exposure distortion map was built as the difference between the positions corrected with the 5th- 
order polynomial (x$_{i}^{c}$, y$_{i}^{c}$), and the positions corrected using only linear transformations
(thus taking into account rigid shifts, rotations and different scale).
For the upper left corner of the camera, the polynomial solution is extrapolated, since no
stars in common with the HST FoV of Ma13 were found. Therefore the solution in that small area might not
be representative of the true distortions.
For all of the stars in common to all of the eight exposures per filter, the difference vectors were averaged 
and then multiplied by 20 to enhance the details.

Several structures in the distortion maps are in common among all of the chips and both the filters.
The amplitude of the distortions in the X-component (corresponding to RA) 
is significantly larger than that in the Y-component (Dec). In fact the 
former spans an interval ranging from $-1.86$ pixels to $4.83$ pixels (mean value of $0.6$ pixels), 
while the latter varies from $-0.82$ pixels to $0.52$ pixels (mean value of $-0.01$ pixels).
The corners of each chip appear to be more distorted.
Since the size of each pixel is that of the HST/ACS, the largest distortions are of the order of 0.2 arcsec.
Similar values were found independently on another GeMS data set (S.Saracino, private communication).
A quite striking common feature is a circular structure roughly located at the centre of each chip 
(which does not move from exposure to exposure) where the X-component distortion changes its sense.
Another remarkable feature is the similarity of the distortions in the four separated chips. Not only
the structure, but also their magnitude is very similar. For all the four chips, minimum, maximum
and mean distortion values in X and Y agree within $\sim0.1$ pixels

\begin{figure*} [ht]
    \begin{center}
    \includegraphics[width=8.5cm]{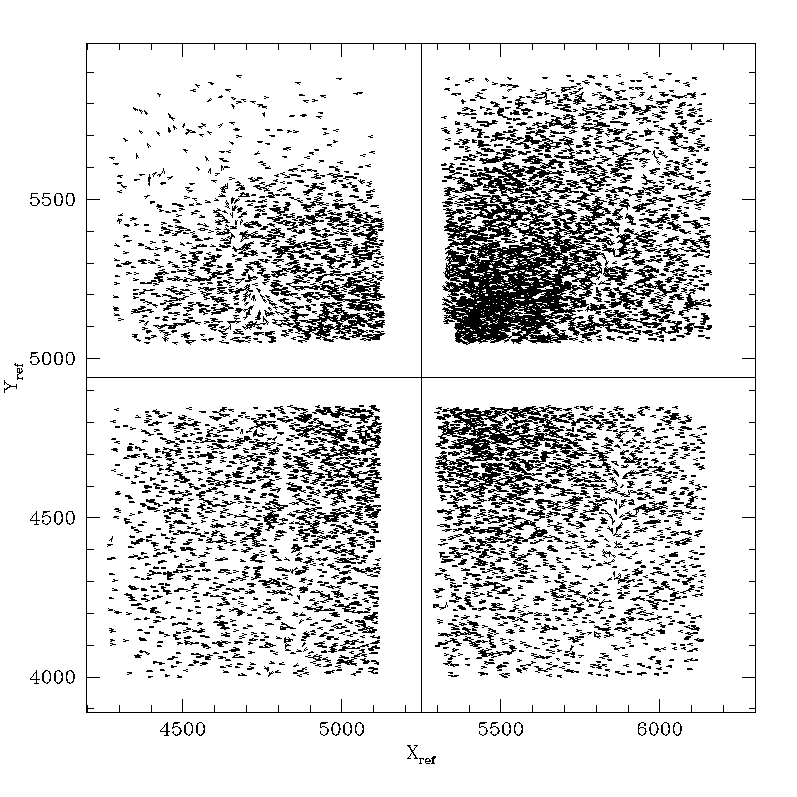}
    \includegraphics[width=8.5cm]{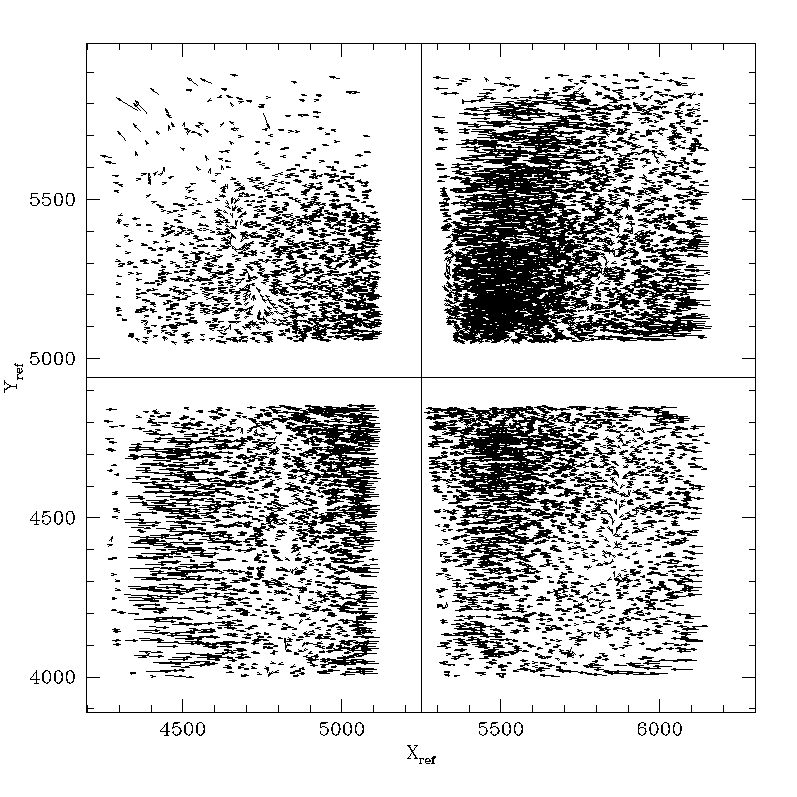}
        \caption{\small Maps of distortion residuals for one of the K-band exposures. The left panel shows the residuals
        magnified by the same factor as that used for the distortion maps (20). Instead, the right panel shows them 
        magnified by 80, to enhance the detail.}
        \label{resi}
    \end{center}
\end{figure*}

Finally, to separate the distortion component that changes from exposure to exposure,
we build residual maps, that is the difference between each single distortion map and 
the average one of the corresponding filter. An example for a K-band exposure is shown in Figure \ref{resi}. In the left-hand 
panel each vector is multiplied by a factor of 20, and the good work made by the average map in describing
the overall behaviour of the single-exposure map is evident. In the right-hand panel, instead, the vectors
are multiplied by a factor of 80, to enhance the residuals. The main structures observed in the 
average maps have disappeared, and only a different pattern survive. The magnitude of the residual distortions
are again very similar among each chip, and ranges from $-1.1$ pixels to $1.3$ pixels (mean value $0.05$ pixels)
in the X component, and from $-0.11$ pixels to $0.07$ pixels (mean value $0.0$ pixels) in the Y component.
Note that by excluding the most external corners, the variation ranges are much smaller, being limited
between $-0.02$ to $0.03$ pixels in both components.
Since from exposure to exposure the only significant difference is the seeing,
we interpret these residual maps as the distortion variations introduced by the varying observing conditions.
This will likely play a significant role in our ability to carry out accurate astrometry and photometry,
and will need to be further studied.

\subsection{Discussion on the distortion maps}

Using previous PM studies with HST we were able to acquire a-priori knowledge to break the PM-distortions 
degeneracy for GeMS observations of the GC NGC6681, and are thus able to model the time varying distortions of each exposure
taken with the MCAO-assisted camera.
Our findings show that the average distortion across the entire FoV amounts to $\sim30$ mas, but it can reach
values as large as $\sim200$ mas. With the use of a fifth order polynomial, we were able to model the distortion to
an accuracy of $\sim1$ mas. A contribution to this term is given by the propagation of the PM error of the stars
used to build the distortion-free reference frame at the GeMS epoch. Since only stars with an error smaller
than 0.03 \masyr were used, and the GeMS data were taken $6.9$ yr after the first HST epoch, the total 
contribution due to PM uncertainty is $\sim0.2$ mas.

Investigations of the internal dynamics of GCs, which might shed light on fundamental topics such as the presence of 
intermediate mass black holes or the formation of GC multiple populations (e.g. \cite{piotto15}), require precisions $<0.1$
mas/yr. Therefore an accurate modelling of such distortions is fundamental to the success of these studies.
The method used throughout this analysis proved efficient in achieving this goal (Massari et al. in prep.) but is limited
to cases for which already measured PMs are available at least for the bright stars in the field. Therefore, 
other solutions still have to be investigated. For example, our group is performing a series of studies 
aimed at anchoring GeMS astrometry to the distortion free reference frame
provided by seeing-limited observations obtained with the FLAMINGOS-2 camera at the Gemini-South telescope.
Looking to the future, MICADO astrometry might also require to test the use of calibration masks or methods that
do not rely on an absolute distortion correction but on a relative calibration.

%

\acknowledgments 
We thank Benoit Neichel, Carmelo Arcidiacono, Jessica Lu and Marc Ammons for the useful discussions on the GeMS distortions. 
Based on observations obtained at the Gemini Observatory and acquired through the Gemini Science Archive.
GF and DM has been supported by the FIRB 2013 (MIUR grant RBFR13J716).

\bibliography{report} 
\bibliographystyle{spiebib} 

\end{document}